\documentstyle[12pt]{article}

\def\be{\begin{equation}}
\def\ee{\end{equation}}
\def\bea{\begin{eqnarray}}
\def\eea{\end{eqnarray}}

\begin{document}

\begin{flushright}
hep-th/9806201
\end{flushright}

\pagestyle{plain} 

\def\e{{\rm e}}
\def\cs{(2\pi\alpha')^2}
\def\CV{{\cal{V}}}
\def\haf{{\frac{1}{2}}}
\def\tr{{\rm Tr}}
\def\goes{\rightarrow}
\def\goal{\alpha'\rightarrow 0}
\def\exp{{\rm exp}}

\begin{center}
{\Large {\bf Feynman Graphs from D-Particle Dynamics}}

\vspace{.2cm}
\small
Amir H. Fatollahi\\
\vspace{.2 cm}
{\it Institute for Studies in Theoretical Physics and Mathematics (IPM),}\\ 
{\it P.O.Box 19395-5531, Tehran, Iran}\\
\vspace{.05 cm}
{\it and}\\
\vspace{.05 cm}
{\it Department of Physics, Sharif University of Technology,\\ P.O.Box
11365-9161, Tehran, Iran}\\ 
\vskip .1 in 
{\sl fath@theory.ipm.ac.ir}
\end{center}
\begin{abstract}
It is argued that quantum propagation of D-particles in the limit 
$\goal$ can represent the "joining-splitting" processes 
of Feynman graphs of a certain field theory in the light-cone frame. So 
basically it provides the possibility to define a field theory by 
its Feynman graphs. The application of this observation to define 
M-theory by an energetic expansion approach is discussed.
\end{abstract}
\vskip .06 in 


Perturbation string theory is defined as sum over different topologies 
of the string world-sheet with weights proportional to the action of string.
The perturbation series can be presented 
as sum over different graphs which describe the "joining-splitting"(JS) 
processes of strings as events in space-time. In fact these graphs are 
those which should be obtained as Feynman graphs of a 
{\it string field theory}. In spite of the absence of a 
string field theory some limits of these graphs, especially those 
produced by small strings ($\goal$) correspond to  
well known particle field theories namely 
various (super) gravity and gauge theories. So basically one expects 
to have a similar interpretation for the particle field theory
graphs as events representing JS processes of particles
in space-time. This expectation will be motivated more when one 
notes that all of the above achievements of string theory to produce 
different field theories is only at a first quantised description 
of strings. 

Recently through the developments of the understanding of M-theory 
as a M(atrix) model \cite{BFSS}, a description of perturbative 
string theory was achieved based on a non-Abelian gauge theory 
living on a cylinder which sometimes can be interpreted as the  
world-sheet of free strings. This description was mentioned 
in \cite{Mot,BaSe} and more concreted in \cite{DVV}.
This picture have been checked in bosonic and super 
string theory in \cite{Fro1} and \cite{Fro2} respectively.
Also in \cite{Wyn} (see also \cite{HV}) it was argued that configurations 
with different lengths of strings and their JS processes 
are corresponded to various backgrounds of world-sheet (gauge) fields 
. So by such description based on gauge theory world-sheet for JS
processes of strings it is tempting to find a similar one 
for D-particles.

It is argued in this letter that different JS processes of D-particles 
have the potential to produce and be corresponded to certain particle 
field theory graphs and their related amplitudes, but in the 
light-cone frame. So it provides the possibility to define a particle 
field theory by its Feynman graphs, those generated by JS processes of  
D-particles. 

Discussions on the question "what field theory?" are 
presented based on M(atrix) theory interpretation of D-particles 
as super-gravitons in the light-cone frame.

D$p$-branes are $p$ dimensional objects which are defined as
(hyper)surfaces which can trap the ends of strings \cite{Po2}. 
One of the most interesting aspects of D-brane dynamics appears in their
{\it coincident limit}. In the case of coinciding $N$ D$p$-branes in a
(super)string theory, their dynamics is captured by a 
dimensionally reduced $U(N)$ (S)YM theory from (9)25+1 to 
$p+1$ dimensions of D$p$-brane world-volume \cite{W,Po2}. 

In case of D-particle $p=0$, the above dynamics reduces to quantum mechanics
of matrices because only time exists in the world-line. The bosonic part
of the corresponding Lagrangian is \cite{BFSS, KPSDF} 
\bea\label{1.1}
L=m_0  \tr\; \biggl(\haf  D_tX_i^2 +\;
\frac{1}{\cs}\;[X_i,X_j]^2\;\biggl) , 
\eea 
where $\frac{1}{2\pi\alpha'}$ and $m_0=(\sqrt{\alpha'}g_s)^{-1}$ are
the string tension and the mass of D-particles, with $g_s$ as string 
coupling. Here $D_t=\partial_t-iA_0$ acts as covariant derivative 
in the 0+1 dimensional gauge theory. 

For $N$ D-particles $X$'s are in adjoint representation of $U(N)$ and have the
usual expansion $X_i=x_{ia}T_a,\;\;a=1,\cdot\cdot\cdot,N^2$. 

Firstly let us search for D-particles in the above Lagrangian:\\
For each direction $i$ there are $N^2$ variables and not $N$ which one
expects for $N$ particles. Although there is 
an ansatz for the equations of motion
which restricts the $T_{(a)}$ basis to its $N$ dimensional Cartan
subalgebra. This ansatz causes vanishing the potential and one
finds the equations of motion for $N$ free particles. In this case the $U(N)$
symmetry is broken to $U(1)^N$ and the interpretation of $N$ remaining
variables as the classical (relative) positions of $N$ particles is
meaningful. The centre of mass of D-particles is represented by the trace
of the $X$ matrices and easily can be seen that the center of mass momentum
is conserved.

In the case of unbroken gauge symmetry, the $N^2-N$ non-Cartan elements have
a stringy interpretation, governing the dynamics of low lying 
oscillations of strings stretched
between D-particles. Although the gauge transformations mix the entries
of matrices and the interpretation of positions 
for D-particles remains obscure \cite{Ba},
 but even in this case the centre of mass is meaningful and the
ambiguity about positions only comes back 
to the relative positions of D-particles.
 
Let us concentrate on the limit $\goal$ 
\footnote{Here this limit
is analogous of the limit $g_s\goes 0$ in \cite{DVV}.}.
In this limit to have a finite energy one has 
\bea\label{1.2}
[X_i,X_j]=0,\;\;\;\forall\;i,j,
\eea
and consequently vanishing the potential term in the action.
So D-particles do not interact and the action reduces to the
action of $N$ free particles 
\footnote{Repeatedly we forget $A_0$ even in path integrals.}
\bea\label{1.3}
S=\int dt \sum_{a=1}^N \haf m_0 \dot{x}_a^2.
\eea
But the above observation fails in the times which D-particles arrive
each other. When two D-particles come very near each other two eigenvalues
of $X_i$ matrices will be equal and this make the possibility that the 
corresponding off-diagonal elements take non-zero values. 
In fact this is the same story of gauge symmetry restoration.
In summary one may deduce that in the limit $\goal$ D-particles
do not interact with each other except for when they coincide.

For two coincident D-particles one may write the corresponding Lagrangian and
Hamiltonian as
\bea\label{1.4}
S=\int dt \bigg(\haf (2m_0) \dot{X}^2 + 
L_{int}(\hat{x}_a,\dot{\hat{x}}_a)\bigg),\;\;\;a=1,2,3,
\eea
\bea\label{1.44}
H=\frac{P^2}{2(2m_0)} + H_{int}(\hat{x}_a,\hat{p}_a),
\eea
which $X$ and $P$ are the position and momentum of the center of mass and
$\hat{x}_a$'s and $\hat{p}_a$'s are the non-Abelian (not only non-Cartan) 
positions and momentums. 
$L_{int}$ and $H_{int}$ are responsible for interactions.


{\it Amplitudes in $\goal$ limit:}\\
Take the probability amplitude corresponded to detecting two
D-particles in positions $x_1$ and $x_2$ at time $t_i$ and 
in $x_3$ and $x_4$ at time $t_f$, presented by path integral as 
\bea\label{1.5}
< x_3, x_4;t_f | x_1,x_2;t_i >=\int \e^{-S}.
\eea
\begin{center}
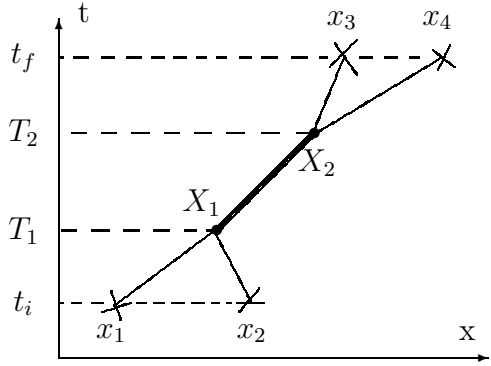
\begin{figure}[t]
\unitlength 1.00mm
\linethickness{0.4pt}
\begin{picture}(110.00,54.33)
\put(53.00,8.33){\line(0,1){43.67}}
\put(53.00,53.33){\vector(0,1){0.2}}
\put(53.00,51.33){\line(0,1){2.00}}
\multiput(60.34,15.33)(0.16,0.12){84}{\line(1,0){0.16}}
\multiput(73.67,25.33)(0.12,-0.23){42}{\line(0,-1){0.23}}
\multiput(73.67,25.00)(0.12,0.12){112}{\line(0,1){0.12}}
\put(87.00,38.33){\line(-1,0){0.33}}
\multiput(86.67,38.33)(-0.12,-0.12){109}{\line(-1,0){0.12}}
\put(73.67,25.33){\line(1,0){0.67}}
\multiput(74.34,25.33)(0.12,0.12){109}{\line(1,0){0.12}}
\multiput(91.00,48.66)(-0.12,-0.29){37}{\line(0,-1){0.29}}
\multiput(86.67,38.00)(0.20,0.12){89}{\line(1,0){0.20}}
\put(87.00,38.33){\circle*{1.33}}
\put(74.00,25.33){\circle*{1.33}}
\multiput(60.00,17.33)(0.11,-0.33){12}{\line(0,-1){0.33}}
\multiput(62.67,15.66)(-0.33,-0.11){12}{\line(-1,0){0.33}}
\multiput(79.67,17.66)(-0.12,-0.14){23}{\line(0,-1){0.14}}
\put(76.67,15.66){\line(1,0){3.67}}
\put(74.34,15.66){\line(-1,0){1.00}}
\put(72.34,15.66){\line(-1,0){1.67}}
\put(69.67,15.66){\line(-1,0){1.67}}
\put(67.00,15.66){\line(-1,0){1.33}}
\put(64.67,15.66){\line(-1,0){1.33}}
\put(59.00,15.66){\line(-1,0){1.66}}
\put(56.00,15.66){\line(-1,0){1.33}}
\put(53.67,15.66){\line(-1,0){0.67}}
\put(53.00,48.33){\line(1,0){0.67}}
\put(55.00,48.33){\line(1,0){2.00}}
\put(58.67,48.33){\line(1,0){1.67}}
\put(62.00,48.33){\line(1,0){2.00}}
\put(65.34,48.33){\line(1,0){1.66}}
\put(68.34,48.33){\line(1,0){1.66}}
\put(71.34,48.33){\line(1,0){1.33}}
\put(74.00,48.33){\line(0,1){0.00}}
\put(74.00,48.33){\line(0,1){0.00}}
\put(74.00,48.33){\line(0,1){0.00}}
\put(74.00,48.33){\line(0,1){0.00}}
\put(74.00,48.33){\line(1,0){1.34}}
\put(77.00,48.33){\line(1,0){1.67}}
\put(80.00,48.33){\line(1,0){2.00}}
\put(84.00,48.33){\line(1,0){2.00}}
\put(87.67,48.33){\line(1,0){1.00}}
\put(93.00,48.33){\line(1,0){1.67}}
\put(96.34,48.33){\line(1,0){1.66}}
\put(99.67,48.33){\line(1,0){1.33}}
\multiput(102.67,49.33)(0.15,-0.12){20}{\line(1,0){0.15}}
\multiput(105.00,50.00)(-0.12,-0.24){14}{\line(0,-1){0.24}}
\multiput(89.67,50.33)(0.12,-0.17){26}{\line(0,-1){0.17}}
\multiput(92.67,50.66)(-0.12,-0.12){31}{\line(-1,0){0.12}}
\put(48.34,48.33){\makebox(0,0)[cc]{$t_f$}}
\put(48.34,15.66){\makebox(0,0)[cc]{$t_i$}}
\put(86.34,38.33){\line(-1,0){1.67}}
\put(82.34,38.33){\line(-1,0){2.34}}
\put(77.67,38.33){\line(-1,0){2.33}}
\put(73.34,38.33){\line(-1,0){2.34}}
\put(69.00,38.33){\line(-1,0){2.00}}
\put(65.00,38.33){\line(-1,0){2.33}}
\put(60.67,38.33){\line(-1,0){2.00}}
\put(56.67,38.33){\line(-1,0){2.00}}
\put(53.34,25.33){\line(1,0){2.00}}
\put(57.34,25.33){\line(1,0){2.33}}
\put(61.67,25.33){\line(1,0){2.00}}
\put(65.67,25.33){\line(1,0){2.00}}
\put(69.67,25.33){\line(1,0){1.67}}
\put(72.67,25.33){\line(1,0){1.33}}
\put(48.34,25.33){\makebox(0,0)[cc]{$T_1$}}
\put(48.34,38.33){\makebox(0,0)[cc]{$T_2$}}
\put(53.00,8.33){\line(1,0){53.00}}
\put(110.00,8.33){\vector(1,0){0.2}}
\put(105.00,8.33){\line(1,0){5.00}}
\put(75.34,15.66){\line(1,0){0.66}}
\put(107.34,11.66){\makebox(0,0)[cc]{x}}
\put(60.00,11.66){\makebox(0,0)[cc]{$x_1$}}
\put(78.67,11.66){\makebox(0,0)[cc]{$x_2$}}
\put(72.00,29.00){\makebox(0,0)[cc]{$X_1$}}
\put(87.34,34.33){\makebox(0,0)[cc]{$X_2$}}
\put(90.67,53.33){\makebox(0,0)[cc]{$x_3$}}
\put(103.34,53.33){\makebox(0,0)[cc]{$x_4$}}
\put(56.34,54.33){\makebox(0,0)[cc]{t}}
\end{picture}
\vspace{-1cm}
\caption{{\it A typical tree path.}}
\end{figure}
\end{center}
\vspace{-.7cm}  
In the limit $\goal$ in that parts of paths which D-particles 
are not coincident, only the diagonal matrices have contribution
to the path integral. This is because of large value of 
action in the exponential. So the action in 
the path integral reduces to the action of two free 
D-particles for non-coincident paths, i.e. $(x_1,x_2)$ till $X_1$ in 
Fig.1. Accordingly one may write, Fig.1 
\footnote{Here as the same which one does in field theory we
have dropped the dis-connected graphs. Also in Fig.1 many other graphs
with different number of loops could be drawn. 
We will come back to them later.},
\bea\label{1.10}
&~&<x_3, x_4;t_f | x_1,x_2;t_i >=\left[\int \e^{-S}\right]_{\goal}=
  \int_{t_i}^{t_f} (dT_1  dT_2 )
  \int_{-\infty}^{\infty} (d^dX_1 d^dX_2 ) \nonumber\\
&~&\times  \bigg(K_{m_0}(X_1,T_1;x_1,t_i) K_{m_0}(X_1,T_1;x_2,t_i)\bigg)
\nonumber\\
&~&\times
\bigg(K_{2m_0}(X_2,T_2;X_1,T_1) 
K_{int}(\hat x_{2a},T_2;\hat x_{1a},T_1)\bigg)
\nonumber\\
&~&\times
 \bigg(K_{m_0}(x_3,t_f;X_2,T_2) K_{m_0}(x_4,t_f;X_2,T_2)\bigg),
\eea
which $K_{m}(y_2,t_2;y_1,t_1)$ is the non-relativistic propagator 
of a free particle with mass $m$ between $(y_1,t_1)$ and $(y_2,t_2)$
, and  $K_{int}$ is the corresponding propagator for
the non-Abelian path integrations. 
In the above relation $\int dT_1 dT_2 dX_1 dX_2$ is for a summation over 
different JS times and points. 

By translating the above to momentum space using 
($E_k=\frac{p_k^2}{2m_0}$, $k=1,2,3,4$)
\bea\label{1.12}
&~&< p_3,p_4;t_f |p_1, p_2;t_i  >\sim 
\e^{i(E_3+E_4)t_f-i(E_1+E_2)t_i}
\nonumber\\
\times\int &~&\prod_{k=1}^4 d^dx_k \e^{i(p_1x_1+p_2x_2-p_3x_3-p_4x_4)}
< x_3,x_4;t_f | x_1, x_2;t_i >,
\eea
and doing integrations one finds (for $t_i=-\infty$ and $t_f=\infty$)
\footnote{We use in $d$ dimensions the representations
$$
K_{m}(y_2,t_2;y_1,t_1)=\theta(t_2-t_1)
\frac{1}{(2\pi)^d} \int d^dp 
\;\exp\bigg(ip\cdot(y_2-y_1)-\frac{ip^2(t_2-t_1)}{2m}\bigg),
$$
$$
K_{int}(\hat x_{2a},T_2;\hat x_{1a},T_1)
=\int d\hat{x}_a \e^{-S_{int}[T_2,T_1]}=
\sum_{n} <\hat{x}_{2a}|n><n|\hat{x}_{1a}> \e^{-iE_n(T_2-T_1)},
$$ 
where $E_n$'s are the eigenvalues of $H_{int}$ of (\ref{1.44}) and 
$\theta(t_2-t_1)$ is the step function.}
\bea\label{1.15}
< p_3,p_4;\infty |p_1, p_2;-\infty  >&\sim&
\delta^{(d)}(p_1+p_2-p_3-p_4)\delta(E_1+E_2-E_3-E_4)
\nonumber\\
&~&\times \lim_{\epsilon \rightarrow 0^+}
\sum_n C_n\; \frac{i}{E-\frac{q^2}{2(2m_0)}-E_n+i\epsilon},
\eea
where $\vec q=\vec p_1 +\vec p_2$ and $E=E_1+E_2$. Now
by recalling the energy-momentum relation in the light-cone frame 
for a particle with mass $M$ 
$$
E\equiv P_+=\frac{{\vec{p}}^2 +M^2}{2P_-}           ,
$$
one sees that the fraction in the sum of (\ref{1.15}) is 
the "light-cone" propagator \cite{Suss} of a particle by identifications
\footnote{In the supersymmetric case of M(atrix) theory there is an easy 
answer about the least value of $E_n$'s, i.e. $E_0$. 
D-particles can make marginal bound states ($E_0=0$)
and so: $M_0=0$. It is an important ingredient when one
wants to identify the above graphs with those ones which come from
supergravity in light-cone frame.}
\bea
P_-&=&2m_0,\nonumber\\
M_n^2&=&4m_0 E_n.
\eea
The first relation learns to us that each D-particle 
has light-cone momentum equal to $m_0$ to have the light-cone 
momentum for two of them $2m_0$ in the time interval $[T_1,T_2]$ in Fig.1.

Now one sees that (\ref{1.15}) is the same expression which one writes 
(in momentum space) as tree diagram contribution to 4-point function of 
a field theory but in the light-cone frame \cite{Suss}, with exchanging masses
as $M_n$'s.

In a more covariant form one may write for (\ref{1.15}) 
\bea\label{1.16}
<< p_3^\mu,p_4^\mu;\infty |p_1^\mu, p_2^\mu;-\infty  >&\sim&
\delta(\vec p_1+\vec p_2-\vec p_3-\vec p_4)
\delta(p_{1+}+p_{2+}-p_{3+}-p_{4+})
\nonumber\\
&~&\times \lim_{\epsilon \rightarrow 0^+}
\sum_n C_n\; \frac{i4m_0}{ q_\mu q^\mu - M_n^2+i\epsilon},
\eea
where $q^\mu=p_1^\mu+p_2^\mu$ and
$$
V^\mu\equiv (\frac{V_++V_-}{\sqrt 2},\frac{V_+-V_-}{\sqrt 2},\vec V),
\;\mu=0,d+1,1,2,...,d,
$$  
$$ 
p_{k+}\equiv \frac{ \vec p_k^2}{2m_0}, \;\;\;\; p_{k-}\equiv m_0, 
\;\;k=1,2,3,4.
$$
The longitudinal momentum conservation trivially is satisfied. Besides
because of conservation of this momentum one can not expect so-called 
$t$-channel processes.

{\it Loop diagrams:}\\
As is apparent loop-ed paths have contributions to the path integral 
(\ref{1.5}). In fact the graph in Fig.1 is the first connected graph. 
One can justify that generically the probability amplitudes associated 
to the loop graphs is the same which one writes in field theory Fig.2. 

There is an important difference between tree and loop graphs related to 
their contributions to the path integral. 
 Because of free moving of D-particles in times which D-particles 
are not coincident there is no classical solution of equations of motion
which can be (topologically) corresponded with a 
graph with loop(s). But it is not
the case for tree diagrams, i.e. there are classical solutions
which are corresponded with tree paths. So the dominant contribution 
to the path integral is due to the tree diagrams; the diagrams which
are corresponded to classical solutions or some deviations around them.
This also indicates a possible well defined perturbative expansion 
for the graphs, because graphs with more vertices have less 
contribution to the amplitudes.
\begin{center}
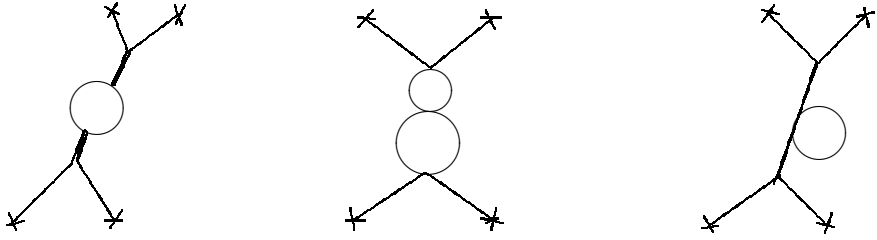
\begin{figure}[t]
\unitlength 1.00mm
\linethickness{0.4pt}
\begin{picture}(122.33,36.33)
\multiput(8.00,7.33)(0.12,0.12){64}{\line(1,0){0.12}}
\multiput(15.67,15.00)(0.12,0.31){14}{\line(0,1){0.31}}
\multiput(17.33,19.33)(0.11,-0.11){3}{\line(1,0){0.11}}
\multiput(17.67,19.00)(-0.11,-0.31){12}{\line(0,-1){0.31}}
\multiput(16.33,15.33)(0.12,-0.19){42}{\line(0,-1){0.19}}
\put(19.00,22.33){\circle{7.09}}
\multiput(21.00,35.33)(0.12,-0.31){17}{\line(0,-1){0.31}}
\multiput(23.00,30.00)(-0.12,-0.27){17}{\line(0,-1){0.27}}
\put(21.00,25.33){\line(1,0){0.33}}
\multiput(21.33,25.33)(0.12,0.31){14}{\line(0,1){0.31}}
\multiput(23.00,29.67)(0.16,0.12){45}{\line(1,0){0.16}}
\multiput(7.33,8.33)(0.11,-0.26){9}{\line(0,-1){0.26}}
\multiput(7.00,6.67)(0.39,0.11){6}{\line(1,0){0.39}}
\multiput(22.33,8.67)(-0.12,-0.17){14}{\line(0,-1){0.17}}
\put(20.00,7.33){\line(1,0){2.33}}
\multiput(21.67,36.33)(-0.11,-0.17){12}{\line(0,-1){0.17}}
\multiput(20.00,35.67)(0.22,-0.11){9}{\line(1,0){0.22}}
\multiput(29.33,36.00)(0.11,-0.30){9}{\line(0,-1){0.30}}
\multiput(30.67,36.00)(-0.11,-0.22){12}{\line(0,-1){0.22}}
\multiput(53.00,7.33)(0.18,0.12){53}{\line(1,0){0.18}}
\multiput(62.67,13.67)(0.17,-0.12){53}{\line(1,0){0.17}}
\put(63.00,17.67){\circle{7.77}}
\put(63.33,24.67){\circle{5.96}}
\multiput(54.33,34.33)(0.16,-0.12){56}{\line(1,0){0.16}}
\multiput(63.33,27.67)(0.15,0.12){56}{\line(1,0){0.15}}
\multiput(52.67,9.00)(0.11,-0.50){6}{\line(0,-1){0.50}}
\put(52.00,7.33){\line(1,0){2.67}}
\multiput(72.00,9.00)(-0.11,-0.50){6}{\line(0,-1){0.50}}
\multiput(70.67,7.00)(0.28,0.11){6}{\line(1,0){0.28}}
\multiput(55.67,35.33)(-0.12,-0.14){17}{\line(0,-1){0.14}}
\multiput(53.67,34.33)(0.78,-0.11){3}{\line(1,0){0.78}}
\multiput(70.67,35.33)(0.11,-0.22){12}{\line(0,-1){0.22}}
\put(72.67,34.33){\line(-1,0){2.67}}
\multiput(73.00,7.00)(-0.50,0.11){6}{\line(-1,0){0.50}}
\multiput(100.33,6.67)(0.17,0.12){56}{\line(1,0){0.17}}
\multiput(109.67,13.33)(0.12,0.36){42}{\line(0,1){0.36}}
\multiput(114.67,28.33)(-0.12,0.12){53}{\line(0,1){0.12}}
\multiput(121.00,34.33)(-0.12,-0.12){53}{\line(-1,0){0.12}}
\multiput(109.67,13.00)(0.12,-0.12){53}{\line(1,0){0.12}}
\multiput(114.67,28.00)(-0.12,-0.33){48}{\line(0,-1){0.33}}
\multiput(23.33,29.67)(-0.12,-0.24){17}{\line(0,-1){0.24}}
\multiput(16.33,15.33)(0.11,0.41){9}{\line(0,1){0.41}}
\multiput(109.00,36.00)(-0.12,-0.17){14}{\line(0,-1){0.17}}
\multiput(107.33,35.33)(0.39,-0.11){6}{\line(1,0){0.39}}
\multiput(121.00,35.67)(0.11,-0.44){6}{\line(0,-1){0.44}}
\multiput(122.33,35.00)(-0.39,-0.11){6}{\line(-1,0){0.39}}
\multiput(99.67,8.00)(0.11,-0.19){12}{\line(0,-1){0.19}}
\multiput(99.33,6.33)(0.78,0.11){3}{\line(1,0){0.78}}
\multiput(116.67,8.33)(-0.11,-0.30){9}{\line(0,-1){0.30}}
\multiput(114.67,7.00)(0.39,-0.11){6}{\line(1,0){0.39}}
\put(115.00,19.00){\circle{6.67}}
\end{picture}
\vspace{-.7cm}
\caption{{\it Loop paths--diagrams.}}
\end{figure}
\end{center}

\vspace{-.7cm}

{\it Field theory of Feynman graphs:}\\
What field theory is corresponded to the above graphs?
Because of the large number kinds of graphs the answer
seems highly hard and may have only a definite answer for some subsets
of graphs. The analogue case in string theory as mentioned in the 
beginning of this letter is the subset of graphs corresponded with the 
limit $\goal$.

A candidate can be guessed by M(atrix) theory approach to
M-theory \cite{BFSS}. M-theory and type IIA string theory both have 
supergravities in their low energy limit, but in different dimensions 
(11 and 10 respectively). Also at present we have a gauge theoretic 
description of stringy "Feynman Graphs" \cite{Mot,BaSe,DVV}. 
Since the 10 dimensional (IIA) supergravity is corresponded 
to stringy "Feynman Graphs" in the limit $\goal$, 
so one may expect to have a similar program for D-particle "Feynman Graphs" 
and 11 dimensional supergravity. So (after adding sufficient
supersymmetry and extracting tensorial structures) 
one should show that a few first 
terms in expansions like (\ref{1.15}) or (\ref{1.16}) are the same which come 
from 11D supergravity in the light-cone frame
\footnote{I feel a close connection between this discussion and the 
formula (2.6) of \cite{del}.}. 
One may summerize this discussion in Fig.3.
\begin{center}
\begin{figure}[t]
\unitlength 1.00mm
\linethickness{0.4pt}
\begin{picture}(140.33,109.00)
\put(33.33,69.67){\framebox(25.00,10.00)[cc]{M-theory}}
\put(33.33,43.33){\framebox(25.00,10.00)[cc]{IIA String}}
\put(108.33,69.67){\framebox(32.00,10.00)[cc]{11D SUGRA}}
\put(108.33,43.33){\framebox(32.00,10.00)[cc]{10D SUGRA}}
\put(40.66,37.66){\rule{9.00\unitlength}{6.33\unitlength}}
\put(40.66,79.00){\rule{9.00\unitlength}{6.33\unitlength}}
\put(108.33,75.00){\vector(1,0){0.2}}
\put(58.33,75.00){\line(1,0){50.00}}
\put(73.33,71.67){\makebox(0,0)[cc]{Low energy}}
\put(108.33,48.66){\vector(1,0){0.2}}
\put(58.33,48.66){\line(1,0){50.00}}
\put(73.33,51.66){\makebox(0,0)[cc]{Low energy}}
\put(124.33,28.33){\line(-1,0){51.00}}
\put(124.00,43.33){\vector(0,1){0.2}}
\put(124.00,28.33){\line(0,1){15.00}}
\put(95.00,98.33){\makebox(0,0)[cc]{Low energy}}
\put(100.00,24.33){\makebox(0,0)[cc]{Low energy ($\goal$)}}
\put(37.33,37.33){\rule{17.00\unitlength}{7.00\unitlength}}
\put(37.33,78.67){\rule{17.00\unitlength}{6.67\unitlength}}
\put(15.00,15.00){\framebox(58.00,23.33)[cc]{}}
\put(44.33,104.00){\makebox(0,0)[cc]{D-particle "Feynman graphs"}}
\put(44.00,91.00){\makebox(0,0)[cc]{from world-line gauge theory}}
\put(44.00,31.67){\makebox(0,0)[cc]{String "Feynman graphs"}}
\put(44.00,20.00){\makebox(0,0)[cc]{from world-sheet gauge theory}}
\put(45.67,53.33){\vector(0,-1){0.2}}
\put(45.67,69.67){\line(0,-1){16.34}}
\put(45.67,61.67){\makebox(0,0)[cc]{compact.}}
\put(124.33,53.33){\vector(0,-1){0.2}}
\put(124.33,69.67){\line(0,-1){16.34}}
\put(124.33,61.67){\makebox(0,0)[cc]{compact.}}
\put(16.33,86.00){\framebox(55.00,11.33)[cc]{}}
\put(17.00,92.00){\line(-1,0){7.00}}
\put(10.00,92.00){\line(0,-1){16.67}}
\put(10.00,54.67){\vector(0,-1){0.2}}
\put(10.00,76.33){\line(0,-1){21.66}}
\put(10.00,56.33){\line(0,-1){29.66}}
\put(10.00,26.67){\line(1,0){5.33}}
\put(10.00,63.33){\makebox(0,0)[cc]{compact.}}
\put(15.00,84.33){\dashbox{2.00}(58.00,24.67)[cc]{}}
\put(73.00,94.33){\line(1,0){2.00}}
\put(77.00,94.33){\line(1,0){2.00}}
\put(81.00,94.33){\line(1,0){2.00}}
\put(85.00,94.33){\line(1,0){2.00}}
\put(89.00,94.33){\line(1,0){2.00}}
\put(93.00,94.33){\line(1,0){2.00}}
\put(97.00,94.33){\line(1,0){2.00}}
\put(101.00,94.33){\line(1,0){2.00}}
\put(105.00,94.33){\line(1,0){2.00}}
\put(109.00,94.33){\line(1,0){2.00}}
\put(113.00,94.33){\line(1,0){2.00}}
\put(117.00,94.33){\line(1,0){2.00}}
\put(121.00,94.33){\line(1,0){2.00}}
\put(123.67,94.33){\line(1,0){0.67}}
\put(124.33,94.33){\line(0,-1){2.00}}
\put(124.33,90.33){\line(0,-1){2.00}}
\put(124.33,86.67){\line(0,-1){2.00}}
\put(124.33,79.67){\vector(0,-1){0.2}}
\put(124.33,83.67){\line(0,-1){4.00}}
\end{picture}
\vspace{-1cm}
\caption{{\it The dashed lines are the expectations motivated by the
results of this letter. The arrow in the left side was known through
M(atrix) theory compactifications [4]}}
\end{figure}
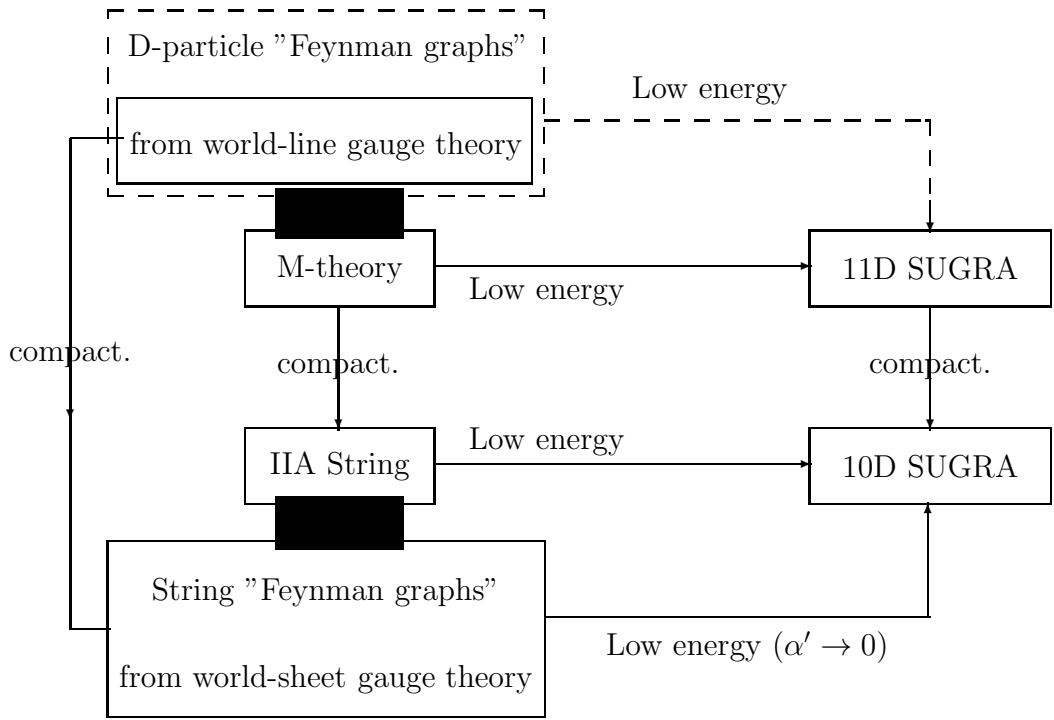
\end{center}
\vspace{-.7cm}
Evidence for the above guess needs more graph-ology.

{\bf Acknowledgement}\\
I am grateful to M.H. Sarmadi who brought my attention 
to gravity as a possible field theory of graphs.

\end{document}